\documentstyle[12pt]{article}

\newcommand{\be}{\begin{equation}}
\newcommand{\en}{\end{equation}}
\newcommand{\bea}{\begin{eqnarray}}
\newcommand{\ena}{\end{eqnarray}}
\newcommand{\vs}[1]{\vspace{#1 mm}}
\newcommand{\ket}[1]{|#1 \rangle}
\newcommand{\bra}[1]{\langle #1 |}
\begin{document}
%
%
%
%
%
%
%
%
%
%
\renewcommand{\thefootnote}{\fnsymbol{footnote}}
\newpage
\setcounter{page}{0}
\pagestyle{empty}
\leftline{KCL-TH-92-4}
\leftline{NI 92007}
\leftline{October 1992}
\vs{20}

\begin{center}
{\LARGE {$W_3$ string scattering}}\\[1cm]
{\large M.D. Freeman}\\[0.3cm]
{\em  Department of Mathematics, King's
College London, Strand,
London WC2R 2LS}\\[0.7cm]
{\large and}\\[0.7cm]
{\large P. West}\\[0.3cm]
{\em  Department of Mathematics, King's
College London, Strand,
London WC2R 2LS}\\
{\em and}\\
{\em The Isaac Newton
Institute for Mathematical Sciences, 20 Clarkson Road,
Cambridge CB3 0EH}
\\[1cm]
\end{center}
\vs{15}
\centerline{ \bf{Abstract}}
The group theoretic method is extended to include fields with a background
charge.
This formalism is used to compute the tree level scattering    for $W_3$
strings.
The scattering amplitudes involve Ising model correlation functions.  A
detailed study of
the four tachyon amplitude shows that the $W_3$ string must possess additional
states in
its spectrum associated with intercept $1/2$ and the energy operator of the
Ising model.
\renewcommand{\thefootnote}{\arabic{footnote}}
\setcounter{footnote}{0}
\newpage
\pagestyle{plain}

%
%
%

In this  letter we  first explain how  to use  the  group  theoretic
approach to  string theory  [1] to  calculate scattering for a
scalar  field  with  a  background  charge.    Such  a  field $\phi$
possesses the energy momentum tensor
$ T = -{1\over 2} (\partial \phi)^2  - Q \partial^2\phi $, which under a
conformal
transformation  $ z \rightarrow f(z) $ induces the transformation
\be
\phi(z) \rightarrow \phi(f(z)) + Q \ln {\partial f \over \partial z}.
\en
The central object in the group theoretic approach is the
N string  vertex, $V^N$,  which corresponds  to a  Riemann surface
with N  marked points  (Koba-Nielsen  coordinates),  $z_i$.    The
vertex is  specified by  relations called unintegrated overlap
equations, which fall into two types; one states how conformal
operators of  a given  weight acting  on different legs of the
vertex are related, while the other relates the transformation
around the  non-trivial homology cycles of such operators when
acting on  the vertex.   To  specify the  overlap relations we
must specify  a coordinate  system $\xi_i$, $i = 1 \ldots N$   in  the
neighbourhood of
each marked point, $z_i$, $i = 1,\ldots ,N$ of the Riemann surface which is
constrained to  have its  origin at its respective point.  The
vertex $V$  is such,  that for  any conformal operator $R(z)$  of
conformal weight d, it obeys the relation
\be
V R^i(\xi^i) (d \xi^i)^d = V R^i(\xi^j) (d \xi^j)^d.
\en

For a  scalar field with a background charge, we take its
corresponding  vertex  to  satisfy  the  unintegrated  overlap
equation
\be
V \phi^i(\xi^i) = V \{ \phi^j(\xi^j) + Q \ln {d\xi^j\over d\xi^i}\}.
\en
Differentiating with  respect to $\xi^i$ we find that V satisfies the
relation
\be
V P^i(\xi) = V \{P^j(\xi^j)  {d \xi^j \over d \xi^i} + i Q {\partial \over
\partial \xi^i}
\ln {d\xi^j\over d\xi^i}\}
\en
for the operator $P(z) = i {d\phi\over d z}$.

The vertex  is completely  determined up to a constant by
these unintegrated  equations; however, it is often quicker to
derive the integrated overlap equations and derive most of the
vertex from these.  These equations are found by considering
the contour  integral  of  $V \Phi P^i(\xi^i)$  around  the  point  $z^i$  and
then
deforming the contour.  The function $\Phi$ is such that it should be
analytic except  that it  may have poles only at the points $z_i$.
As the  contour approaches  another Koba-Nielson  point $z_j$,  we
must change  to the  coordinates $\xi^j$ associated  with this  point.
The inhomogeneous  term in  the transformation  of $\phi$ of equation (1)
does  not lead  to any  singularities and so contribute at
this new  point; however, it does provide a contribution as we
change coordinates  to include  the point  at infinity,  whose
local coordinates  we denote  by  u.    This  latter  step  is
necessary even  if we  have only one leg, since we require two
coordinate patches  to cover  the Riemann  surface.   The  net
result of these manoeuvers is the equation
\be
\sum_{j=1}^N V \oint d \xi^j P^j(\xi^j) f = \oint_0 d u V 2 i Q {f \over u}.
\en
Taking $\Phi$ to  be a  constant we  find  the  well  known  momentum
conservation condition in the presence of a background charge,
namely
\be
\sum_{j=1}^N \alpha_0^j = 2 i Q.
\en
Taking $f\Phi= 1/(\xi^i)^n$ for $n > 0$ we   find  no   contribution  at
infinity  and  as  a
consequence the $\alpha_n^i\alpha_m^j$ and $\alpha_n^i\alpha_0^j$  terms  in
the
vertex are the same as the
case for when the background charge is zero.  To determine the
$\alpha_0^i\alpha_0^j$ terms we must  examine the  unintegrated  overlap  of
equation  (3).
Following the  discussion of  section (3) of reference [2], we
find that for tree scattering the vertex is of the form
\bea
V &=& \{ \prod_{i=1}^N \bra0\} \exp \{-\sum_{1\le i< j\le N}
\{\sum_{n,m =1}^\infty\cr
&& {\alpha^{\mu (i)}_n \over \sqrt n}{(n+m-1)! (-1)^m\over (m-1)!
(n-1)!(z_j-z_i)^{n+m}}
{\alpha^{(j)}_{\mu m} \over \sqrt m}\cr
&& +
\sum_{n=1}^\infty
\{{\alpha^{\mu (i)}_n \over \sqrt n} {\alpha^{(j)}_{\mu 0} \over (z_j-z_i)^n}
+ {\alpha^{\mu (j)}_n \over \sqrt n} {\alpha^{(i)}_{\mu 0} \over (z_i-z_j)^n}\}
 + \alpha^{(i)\mu}_0 \alpha^{(j)}_{0 \mu} ln (z_j-z_i)\}\}\cr
&&\delta(\sum\alpha^\mu_0 + 2 i Q)
\ena
for  the simplest  possible relation between
the coordinate  patches, namely $\xi^i = \xi^j + z^j - z^i$.   This result is
in agreement
with reference [3].  In fact for this cycling transformation the additional
term in
the overlap of equation (3) vanishes for transitions between the patches
for the marked points.  However, one can verify for any $sl(2,R)$ transition
functions that the vertex, when written in a form in which it contains in
the exponential no terms bilinear in the same species of oscillator,
has a $Q$-dependence only in the momentum conservation
$\delta$-function.  The vertex for any transition functions can then be easily
read off from past papers.
To verify this statement we must redo the calculation of reference [2]
to find the $Q$-dependence we
must include the term $2 i Q$ whenever we use momentum conservation
and take into account the additional terms in the overlap equation.
In fact momentum conservation is used twice; to reprocess the second
term of equation (3.9) and
when cancelling the last two terms of equation (3.10).  This leads to an
additional term $(-i)(-2 i Q) \ln(c^{ij} - a^{ij} - \xi^i)$ in the notation of
that
paper, except that we take $z=\xi^i$ and include a factor of $-i$ to gain a
more usual definition of $\phi$.  This term is precisely cancelled by the
additional $-Q \ln d\xi^j/d \xi^i$ term in the overlap.

Loops  are   also  easily   incorporated,  but   now  the
transformations around the non-trivial homology cycles reflect
the presence of the additional term in the transformation of $\phi$.
For  example,  for  the  Schottkey  representation,  we  would
implement the equation
\be
V P^i(\xi^i) = V P^j(P_n^j \xi^j) {d (P_n^j \xi^j) \over d \xi^j}
+ i Q {\partial\over\partial \xi^i} \ln {d (P_n^j \xi^j) \over d \xi^j}
\en
for the  transformation $P_n$  between the  two circles associated
with the $n$'th  loop.

The changes  of the  moduli dependence  follow the  usual
path, but  now we  must take  into account  equation (1).   We
illustrate it  for the  simplest  case,  namely  changing  the
position of  a Koba-Nielsen coordinate.  Under $z\rightarrow  z + a$, which
is induced  by $e^{a L_{-1}}$, equation (1) takes the
form $\phi(z) \rightarrow \phi(z+a)$.  Acting with $e^{a L_{-1}}$
on equation  (3), we find an integrated overlap for the vertex
$V e^{a L_{-1}}$ ,  which we  readily interpret as being the old vertex, but
with $z^i \rightarrow z^i + a$.  Infinitesemally, we have the usual result
\be
{\partial V \over \partial z^i} = V L_{-1}.
\en
The energy momentum tensor $T(z)$ includes the background  charge, but
obeys the same conformal transformation law as the more usual case, and
as such it will obey the usual overlap [1]
\be
V \sum_{j=1}^N \oint d \xi^j T(\xi^j) \phi =  0.
\en
We may have to, as usual, add constants [1], although this is not the case
for tree level vertices.

In the group theoretic method the scattering amplitude is
given by integrating the vertex over the moduli space weighted
with a function of the moduli which is determined by demanding
that zero  norm physical  states decouple.   Let us illustrate
the procedure  for a  D scalar string,$\mu = 0,1,\ldots , D-1$, which has a
background
charge, $\alpha^\mu$,  adjusted so  that it  is critical,  i.e.
$\alpha\cdot\alpha = (26-D)/12$.
The
physical states  of such  a string  satisfy $(L_n - \delta_{n,0})|\psi\rangle =
0, n \ge 0$,
and it
has a
level 1  null state, i.e.  $L_{-1} |\psi\rangle$
where $L_n|\Omega\rangle = 0, n \ge 0$.
The N string scattering amplitude is given by
\be
\oint \prod_{i=1}^N d z^i V(z^i) f
\en
Demanding that  the above  null states  decouple  implies
that  ${\partial f\over \partial z^i}= 0$ or $f$ is a constant.

The formalism developed in this paper can also be applied to $W_3$ string
scattering.  A $W_3$ string was constructed [6,7] by utilizing the Miura
transformation
and obeying the consistency conditions arising from the BRST charge[8].  This
can be
extended, by generalizing the Miura transformation, to have a $W_3$ string for
any number of
scalar fields [7].  It has been noticed as a phenomenological observation that
the central
charges arising in $W_N$ strings  are related to those of the minimal
models[6,7].
It has been found [4] by explicitly solving the physical state conditions at
low levels, that
the $W_3$ string theory has one of the oscillators suppressed and that
physical,
positive definite norm,
states are contained in a subspace of the original oscillator space that obeys
$L_n|\psi\rangle = 0, n\ge1$ and $(L_0 - a_{eff})|\psi\rangle = 0$ for $a_{eff}
= 1$ or $15/16$.
In these equations the $L_n$'s are the Virasoro operators for $D+1$ scalar
fields
( the original
$W_3$ string having $D+2$ scalar fields) with a background charge tuned to give
a
central charge
of  $51/2$.  The lowest level null states in the $a_{eff} = 1$
sector is $L_{-1}\ket\psi$, where
$L_n\ket\psi = 0$ for $n\ge0$, and so scattering in this sector is given by
$\int\prod_i d z_i V^N f$.  By applying our previous argument we find
that $\partial_i f = 0$
for $i=1, \ldots, N$.

The lowest null state in the $a_{eff}=15/16$ sector, however, is given by
$(L_{-2}  + 4/3 L_{-1}^2) \ket{\Omega'}$, where $L_n \ket{\Omega'} = 0$ for
$n\ge 1$ and
$L_0 \ket{\Omega'} = -17/16 \ket{\Omega'}$.  The scattering is given by
$\int\prod_i d z_i V^N f$; applying the above null state on leg $j$ and
physical states
$\ket{\chi}_i$ on the other legs we find that physical null states decouple if
\be
\int\prod_i d z_i f V^N  (L_{-2}  + 4/3 L_{-1}^2) \prod \ket{\chi}_i
\ket{\Omega'}_j = 0.
\en
Using equation (9) with $f=1/\xi^j$ we can replace $L_{-2}^j$ by $L_n, n\ge-1$,
on the other
legs.  Carrying out this, using equation (7) and the physical state conditions
we find that
\be
\int\prod_i d z_i f \{ {4\over 3}{\partial^2 V\over\partial z^j{}^2} +
\{ {\partial V\over \partial z^i} { 1\over (z^j-z^i)} - {17\over 16}
{L_0^i\over (z^j - z^i)}\}\} \prod \ket{\chi_i}\ket{\Omega'}_j = 0.
\en
Integrating by parts we conclude that the measure $f$ satisfies
\be
\{ {4\over 3}{\partial^2 f \over \partial z^j{}^2} -\sum ({\partial
f\over\partial z^j}
{1\over (z^j - z^i)} + {1\over 16} {f\over (z^j-z^i)^2})\} = 0.
\en
This equation for $f$, however, is none other than the equation obeyed by
the correlation function for $N$ primary fields of conformal   weight $1/16$
of the Ising model.  In fact the measure in the $a_{eff} = 1$ sector can also
be given such an interpretation since the correlation function for $N$
primary fields of weight $h=0$ is independent of $z_i$.  The correlation
function for four weight $1/16$ Ising states was found in reference [5].  Thus
we find an interesting connection between the $W_3$ string scattering and
the Ising model correlation functions.

We could also consider the scattering of strings from the two sectors; let
$i=1,\ldots,M$ be from the $a_{eff}=1$ sector and $i=M+1,\ldots,M+N$ be from
the $a_{eff}= 15/16$ sector.  The vertex satisfies the overlaps of
equations (3) and (4) for any $i,j \in [1,\ldots,M+N]$ as well as the
integrated overlap of equation (5).   As a result the vertex is of the same
form as equation (5) but with the sums over $i = 1,\ldots, M+N$.  It is
straightforward to extend the above arguments for the decoupling of null
states to show that the corresponding measure $f$ obeys the equation
for the Ising correlation function for $N$ fields of weight $1/16$ and
$M$ fields of weight $0$.  We could rewrite the vertex so as to absorb the
measure
by including the
Ising operators $\Xi = \{1,\sigma\}$, where 1 is the identity operator and
$\sigma$ the weight $1/16$ field, in the scattering vertex:
\be
V' = V \prod_{i=1}^N \Xi^i(z_i),
\en
where $\Xi^i$ is valued according to whether the external state has $a_{eff}=1$
or $15/16$
respectively.  With such a vertex the measure can be taken to be one.  One
could also
work in the more standard, but less powerful, conformal field  theory
approach to string theory.
The corresponding vertex operators can be extracted from the three-vertex by
putting on one
leg the appropriate external state and identifying the oscillators of the
remaining  two legs.
The
tachyon vertex operators are $e^{i p\cdot\phi}$, $p^2 = 2$ and
$\sigma e^{i p\cdot\phi}$, $p^2 = 15/16$ in the two sectors respectively.

It will prove instructive to  work out in detail the scattering of four
tachyons in the
$a_{eff}=1/2$ sector.  Putting   four such states on the vertex of equation (5)
we  find, after
choosing $z_1=\infty, z_2 = 1, z_3 = x$ and $z_4 = 0$ the result
\be
F(s,t) = \int_0^1 d x x^{-\alpha' s - 15/8} (1-x)^{-\alpha' t-15/8} f(x)
z_1^{1/8}.
\en
The function $f(x)$ is the four point $\sigma$ Ising correlation function which
is of the   form [5]
\be
f(x) = [ (z_1 - z_3)(z_2 - z_4)]^{-1/8} Y(x),
\en
where
\be
x={(z_1-z_2)(z_3-z_4)\over (z_1-z_3)(z_2-z_4)}
\en
and
\be
Y(x) = {1\over[x(1-x)]^{1/8}} (a \cos\theta + b \sin\theta),
\en
with $x=\sin^2\theta$.  The two constants $a$ and $b$ correspond to the
existence of   two
solutions to the differential equation (13).  Taking the above choice of $z$'s
we find that
\be
F(s,t) = \int_0^1 d x x^{-\alpha's-2} (1-x)^{-\alpha' t-2}(a \cos\theta + b
\sin\theta),
\en
which is finite as $z_1 \rightarrow \infty$.  In the above we have introduced
the slope
$\alpha'$,
which is usually taken to be $1/2$ for the open string.

The values of the constants $a$ and $b$ are to be chosen by a physical
requirement,
which in
our case is crossing.  The open string amplitude $T^{(4)}(p_i)$ is as usual the
sum of three
terms
\be
T^{(4)}(p_i) = F(s,t) + F(t,u) + F(u,s).
\en
Crossing for four identical particles means that $T^{(4)}(p_i)$ should be
symmetric under the
exchange of any two legs or equivalently momenta.  This means for example  that
it should be
symmetric under $s\leftrightarrow t$.  This follows provided that $F$ itself is
a symmetric
function of its arguments.  This property is in turn guaranteed if the
integrand is symmetric under
$x \leftrightarrow 1-x$, while at the same time we interchange $s$ and $t$, or
$p_2$ and $p_4$.
The transformation $x\rightarrow 1-x$ can be written as $\theta \rightarrow
\pi/2 - \theta$,
whereupon it is  obvious that we  should take
\bea
F(s,t) &=& \int_0^1 d x x^{-\alpha's-2} (1-x)^{-\alpha' t-2}
\cos(\theta/2-\pi/8)\cr
&=& {1\over\sqrt 2}\int_0^1 d x x^{-\alpha's-2} (1-x)^{-\alpha' t-2}\{
\cos\pi/8 \sqrt{1+\sqrt{1+x}} + \sin\pi/8 \sqrt{1-\sqrt{1-x}}\}.\nonumber
\ena

Having found the expression for four tachyon scattering we can examine the
particles
exchanged in a given channel.  Let us consider the $s$-channel.  Expanding the
factor
$(1-x)^{-\alpha't-2}$ we find that
\bea
F(s,t)&=&\sum_{n=0}^\infty \sum_{p=0}^\infty  \cos{\pi\over8} \sqrt 2 a_p
{(\alpha't+2)(\alpha't +3)\ldots,\alpha't + n + 1)\over(-\alpha's - 1+p+n)}\cr
&+&\sum_{n=0}^\infty \sum_{p=0}^\infty  \sin{\pi\over8}  {b_p \over \sqrt 2}
{(\alpha't+2)(\alpha't +3)\ldots,\alpha't + n + 1)\over(-\alpha's - 1+p+n+
1/2)},
\ena
where
\be
\sqrt{1+\sqrt{1+x}} = \sqrt 2 \sum_{p=0}^\infty a_p x^p
\en
and
\be
\sqrt{1-\sqrt{1-x}} = \sqrt{x\over2} \sum_{p=0}^\infty b_p x^p.
\en
The states exchanged in the first and second terms have masses which satisfy
$\alpha' m^2 = p + n - 1$ and $\alpha' m^2 = p + n - 1/2$ respectively.  While
the former
states are contained in the $a_{eff} = 1$ sector the latter are in neither the
$a_{eff}=1$
nor the
$a_{eff}=15/16$ sector.  Consequently we find that the $W_3$ string requires
the existence of additional states in its spectrum .    These states correspond
to an intercept
of $1/2$, and so it
is very natural, given the above pattern, to associate them with the weight
$1-1/2=1/2$ operator
$\epsilon$ of the Ising model.  As such, all the operators of the Ising model
appear in the
$W_3$ stringl the tachyon in the new
sector should correspond to the vertex operator $\epsilon(z) e^{i k\cdot\phi}$,
with $k^2 = 1$.
The existence of these new intermediate states is to be expected in view of the
fusion rule
$\sigma \sigma = 1 + \epsilon$.

We now repeat the above discussion for the closed $W_3$ string.  The closed
string overlaps
are the same as those of equation (3), except that now $\phi$
involves the left and right oscillators.
Tracing through the argument following this equation we conclude that the
closed string
vertex is
the product of two open string vertices, one containing left oscillators and
the other  right
oscillators, with a zero mode term in common with appropriate normalization.
The decoupling of null states implies that the measure$f$ satisfies the
appropriate Ising
model equation, and so for  four tachyons in the $a_{eff}=15/16$ sectors for
left and
right it is of
the form
\bea
&&f(z_i,\bar z_i) = [(z_1-z_3)(z_2-z_4)(\bar z_1 - \bar z_3)(\bar z_2 - \bar
z_4)]^{-1/8}
[x \bar x (1-x)(1-\bar x)]^{-1/8}\cr
&&\{u_{11} \cos\theta/2 \cos \bar\theta/2 + u_{12} \cos\theta/2 \sin
\bar\theta/2 +
u_{21} \sin\theta/2 \cos \bar\theta/2 + u_{22} \sin\theta/2 \sin \bar\theta/2
\},\nonumber
\ena
where $\bar x$ is the cross-ratio of equation (16) but with $\bar z$'s instead
of $z$'s, the
$u_{ij}$'s
are constants and $x=\sin^2 \theta$, $\bar x = \sin^2 \bar\theta$.

The amplitude for four tachyon closed string scattering is
\be
T^{(4)}(p_i) =
\int d^2 z_3 |z_1-z_2|^2 |z_1-z_4|^2 |z_2-z_4|^2 \prod_{i<j} |z_i-z_j|^{2
\alpha' k_i
\cdot k_j} f(z_i,\bar z_i).
\en
Taking $z_1=\infty$, $z_2 = 1$, $z_3 = x$ and $z_4 = 0$, this amplitude is
finite and becomes
\be
T^{(4)}(p_i) = \int d^2 x |x|^{-\alpha' s -4} |1-x|^{-\alpha' t -4}
\{ u_{11} \cos\theta/2 \cos \bar\theta/2
+ \ldots u_{22} \sin\theta/2 \sin \bar\theta/2\}.
\en

The constants are again determined by requiring crossing.  However, for the
closed string we
only have one term and so in addition to the symmetry arising from exchanging
particles 2 and 4
which induces $s\leftrightarrow t$, $x \leftrightarrow 1-x$, $\bar x
\leftrightarrow 1-\bar x$ we have
that arising from exchanging particles 4 and 1 which induces $s\leftrightarrow
u$,
$x \leftrightarrow 1/x$ and $\bar x \leftrightarrow 1/\bar x$. There is only
one choice of the
$u_{ij}$'s, namely $u_{11} = u_{22}$ and $u_{12} = u_{21} = 0$, and so the
final amplitude is
\be
T^{(4)}(p_i) = \int d^2 x |x|^{-\alpha' s -4} |1-x|^{-\alpha' t -4}  \cos
({\theta - \bar \theta \over 2}).
\en
The last part of this expression is  none other than the usual Ising result.
It is clear that if we
factorize this amplitude we will find the closed string  analogue of the above
new states.

It would be interesting to confirm the above  scattering   results by working
in the full Fock
space including all the oscillators in the $W_3$ string and not just those that
occur in the
states in the $a_{eff} = 1$ and $15/16$ sectors.  The starting point in this
calculation is
the vertex $\hat V$ which obeys the overlaps of equation (1.3) for the $D+2$
string coordinates.
This vertex is then given by equation (1.7) with the appropriate background
charges
encoded in the $\delta$-function, and it obeys in addition to the T overlap of
equation (1.10)
an overlap for $W$ of the form
\be
\hat V \sum_j \oint d \xi^j W^j(\xi^j)\Phi = 0,
\en
where $\Phi$ is now a second rank tensor.  There are, on the sphere,
5 such analytic tensors which
we may take to be $(\xi^j + z^j)^n$ for $n=0,1,\ldots,4$.  To find the
scattering we
must take account of the $W$ moduli.  This would be best achieved by
incorporating a
geometric understanding of the $W$ algebra analogous to the use of superspace
for the
super Virasoro algebra.  However, one could also proceed using the method of
the third paper
of reference [1], which regarded the vertex as an induced representation.  Its
isotropy group
is generated by the overlap identitiesfor the generators, and two vertices
which differ by
transformations that annihilate on-shell states are considered equivalent.
Hence for an $N$-string
tree level vertex we have for the Virasoro part of the algebra one modulus
(i.e. $L_{-1}$)
associated with each string, subject to 3 identities corresponding to 3
analytic vector
fields, giving in all $N-3$ moduli.  For the $W$ part of the algebra we have 2
moduli
(i.e. $W_{-2}$ and $W_{-1}$) associated with each string, subject to the 5
identities
corresponding to the 5 analytic second-rank tensors.  This gives in all $2N -5$
$W$ moduli.

For a genus $g$ surface ($g\ge2$) with $N$ strings we find, by the same
argument,
$N+3(g-1)$ Virasoro moduli and $2N+5(g-1)$ $W$ moduli.

We can regard the vertex $\hat V$ as being evaluated at zero moduli and we can
introduce by a boost
the $W$ moduli by the action of $W_{-n}$'s on the vertex.  While this is
straightforward
for an infinitessimal $W$ moduli change, i.e. for tree level scattering
\be
V(z_j)(1 + \sum{}'(w_{2j} W_{-2}^j + w_{1j} W_{-1}^j)) = V(z_j,w_{1j},w_{2j}),
\en
where the prime on the sums indicates the absence of 5 terms, a difficulty can
occur for
finite $W$ moduli as one must know how to move from the $W$ algebra to the
$W$ group.  It would be most interesting to confirm the presence of the Ising
correlation
functions from this viewpoint.

In this paper we have shown that the ``group theoretic method'' naturally
encompasses the possibility of fields with a background charge.  This formalism
was
then used to compute the scattering in $W_3$ string theory.  The scattering
amplitudes
were found to involve Ising model correlation functions containing the identity
and spin
operator.  Examining the intermediate states in the four tachyon amplitude
showed that
the $W_3$ string must possess new states in its spectrum associated with a
$1/2$
intercept and the energy operator of the Ising model.  This connection between
the Ising
model and the $W_3$ string will extend to higher minimal models and $W_N$
strings.
It would be interesting to find  the fundamental  reason why the Ising model
plays
such  a crucial role in the $W_3$ string.  A more complete treatment of the
$W_3$ string from a path integral viewpoint may be instructive in this respect.

Note added

While this work was being  written up it was found that the required additional
states in
the $W_3$ string were indeed contained in the cohomology of $Q$, but in
non-standard
ghost-number sectors \cite{LPW}

Acknowledgement

One of the authors (PW) would like to thank J. Cardy, B. Nilsson and C. Pope
for discussions.
MF would like to thank the SERC for financial support.

%

\end{document}